\documentclass[prl,preprint,superscriptaddress]{revtex4-1}
\usepackage{setspace}
\usepackage{amsfonts,amsmath,amssymb}
\usepackage{txfonts}
\usepackage{graphics}
\usepackage{bm}
\usepackage{verbatim}
\usepackage{hyperref}
\usepackage{units}
\usepackage[T1]{fontenc}
\usepackage[latin9]{inputenc}
\usepackage{amsbsy}
\usepackage{esint}
 
\begin{document}

\title{Geometric integration of the Vlasov-Maxwell system with a variational particle-in-cell scheme}

\author{J.~Squire}
\affiliation{Plasma Physics Laboratory, Princeton University, Princeton, New Jersey 08543, USA}
\author{H.~Qin}
\affiliation{Plasma Physics Laboratory, Princeton University, Princeton, New Jersey 08543, USA}
\affiliation{Dept.~of Modern Physics, University of Science and Technology of China, Hefei, Anhui 230026, China}
\author{W.~M.~Tang}
\affiliation{Plasma Physics Laboratory, Princeton University, Princeton, New Jersey 08543, USA}

\begin{abstract}
A fully variational, unstructured, electromagnetic particle-in-cell integrator is developed for 
integration of the Vlasov-Maxwell equations. Using the formalism of Discrete Exterior Calculus \cite{Desbrun:2005p5846}, 
the field solver, interpolation scheme and particle advance algorithm are derived 
through minimization of a single discrete field theory action. As a consequence of 
ensuring that the action is invariant under discrete electromagnetic gauge transformations, 
the integrator exactly conserves Gauss's law.
\end{abstract}

\pacs{52.65.Rr, 52.65.Ff, 52.25.Dg}

%\date{\today}

\maketitle

Particle in cell (PIC) codes have been a crucial tool in understanding
complex plasma dynamics through solution of the Vlasov-Maxwell equations.
The underlying idea of PIC is to advance electromagnetic fields on
a fixed grid, while individual quasiparticles are tracked in continuous
space. This is realized by interpolating fields to particle positions,
advancing positions and velocities in time, then interpolating charge
densities and currents from new particle positions back to the fixed
grid. As modern supercomputers move into the exaflop ($10^{18}$ floating
point operations per second) regime and beyond, PIC codes are increasingly
being used for simulations of larger and more complex systems \cite{RRosner_Computing2010}. To
be able to rely upon the fidelity of simulation results and thus fully
utilize computational resources, it is critical that algorithms have
good long time conservation properties. This is the underlying idea
behind geometric integrators: integrators designed to respect 
geometric principles of the underlying physical system being studied,
thereby reducing spurious numerical effects damaging to the fidelity of simulations. 
In the past decade there has been rapid development
of these techniques, both for time discretization, with variational
integrators \cite{springerlink:10.1007/BF01077598,Marsden:2001varint,Qin:2008p5812},
and for spatial discretization, with Discrete Exterior Calculus (DEC)
and mimetic finite elements \cite{Hyman-2005-principles,Desbrun:2005p5846,Bossavit:1998:Book}.

In this communication we use the ideas of discrete exterior calculus (DEC)
and variational integrators to formulate a geometric PIC scheme that 
conserves a space-time multi-symplectic structure \cite{Marsden:2001varint}. 
While symplectic particle pushing algorithms and multi-symplectic electromagnetic field solvers exist, 
coupling two such schemes does not guarantee multi-symplecticity of the PIC algorithm as a whole. 
Our method is to devise a single \emph{space-time} discrete Lagrangian, 
then use the principle of least action to derive the entire PIC scheme. 
This approach is motivated by the success of numerous integrators
of this type for other field theories. These include integrators for
continuum mechanics \cite{Lew2003_AVIs}, electromagnetism \cite{Stern:2008p5799},
incompressible fluids \cite{Pavlov2011443}, and more complex fluids,
such as ideal magnetohydrodynamics \cite{Gawlik:2010p5814}.
In all cases, the algorithms have very good long time energy conservation
as well as other desirable properties.

In addition to multi-symplecticity, discrete current conservation
$\partial_{t}\rho+\nabla \cdot \bm{J}=0$, is a natural property of the variational formulation: 
it is a direct consequence of discrete electromagnetic gauge invariance of the discrete action. 
Current conservation ensures Gauss's law, $\nabla \cdot \bm{E}=\rho/\epsilon_{0}$, remains satisfied at all times. 
This is important both from a physics
and computational standpoint, since Gauss's law is non-local and can
be difficult to solve efficiently on modern, massively parallel computing
systems. Understanding current conservation in terms of discrete gauge 
invariance could be crucial in the future design of geometrical PIC schemes 
for more complicated field theories, for instance gyrokinetics \cite{Qin:2007p5801}. 

The classical action for a collection of particles interacting with
a self-generated electromagnetic field is, \begin{equation}
\mathcal{S}=-\frac{1}{2}\int_{x}dA\wedge\star dA+\int\sum_{p}\left(q_{p}A+p\right)\mid_{\bm{x}_{p}\left(t\right)}.\label{eq:Continuous action}\end{equation}
Here $A$ and $p$ are 1-forms on 4-D space-time, respectively the
4-vector potential of the field and the particle momentum 1-form,
and $q_{p}$ is the particle charge. $\int_{x}$ denotes integration
over space-time and $\sum_{p}$ denotes the sum over all particles
with $A$ and $p$ evaluated at particle positions, $\bm{x}_{p}\left(t\right).$
The exterior derivative, hodge star and wedge product are all operating
in 4-D space-time. In terms of fields $-\frac{1}{2}dA\wedge\star dA$
is simply $E^{2}-B^{2},$ where we have chosen the geometric notation for
the sake of clarity in the discretization of the action principle.
Throughout this article natural units are used with $c=\varepsilon_{0}=1$. 

In the non-relativistic limit,
\begin{equation}
\left(A+p\right)\mid_{\bm{x}_{p}\left(t\right)}=q_{p}\bm{A}\left(\bm{x}_{p}\right) \cdot d\bm{x}-q_{p}\phi\left(\bm{x}_{p}\right)dt+m_{p}\bm{v}_{p} \cdot d\bm{x}-\frac{1}{2}m_{p}v_{p}^{2}dt,\label{eq:Non rel particle 1 form}
\end{equation}
with $\bm{A}$ and $\phi$ the usual electromagnetic potentials
and $m_{p}$ particle mass. In this limit, the action, Eq.~\eqref{eq:Continuous action},
is simply the mixed Eulerian-Lagrangian action principle of Low expressed in geometric notation \cite{Low:1958p5805,Qin:2007p5801},
with the distribution function $\sum_{p}\delta\left(\bm{x}-\bm{x}_{p}\right)\delta\left(\bm{v}-\bm{v}_{p}\right)$.
The equations of motion for the system are
\begin{eqnarray}
 &  & \bm{\dot{v}}_{p}=\frac{q_{p}}{m_{p}}\left[\left(-\nabla\phi-\frac{\partial\bm{A}}{\partial t}\right)\mid_{\bm{x}_{p}}+\bm{v}_{p}\times\left(\nabla\times\bm{A}\right)\mid_{\bm{x}_{p}}\right],\nonumber \\
 &  & \dot{\bm{x}}_{p}=\bm{v}_{p},\nonumber \\
 &  & d\star dA=\mathcal{J}.\label{eq:Continuous eqns}
 \end{eqnarray}
 Here, $\mathcal{J}=\star\left(\bm{J} \cdot d\bm{x}-\rho \, dt\right)$, with $\bm{J}$  the current density $\sum_{p}q_{p}\bm{v}_{p}\delta\left(\bm{x}-\bm{x}_{p}\right)$
and $\rho$  the charge density $\sum_{p}q_{p}\delta\left(\bm{x}-\bm{x}_{p}\right).$ 

Eqs.~\eqref{eq:Continuous eqns} are gauge invariant, 
meaning an exact 1-form, $df$, can be added to $A$ without
changing the dynamics. This follows directly from symmetry of the
action, Eq.~\eqref{eq:Continuous action}, under the transformation
$A\rightarrow A+df$. The symmetry leads
to the conserved quantity $d_{s}D-\rho$, 
($d_{s}D$ is the divergence of the electric displacement), which is
simply Gauss's law \cite{Stern:2008p5799}. This principle, that gauge invariance of the action
will lead to equations that conserve Gauss's law, is critical
for our discretization of the problem. Note that this is equivalent
to current conservation, $d\mathcal{J}=0$ ($\partial_{t} \,\rho+\nabla \cdot \bm{J}=0$
in standard notation).

\begin{figure}

\begin{centering}
\includegraphics{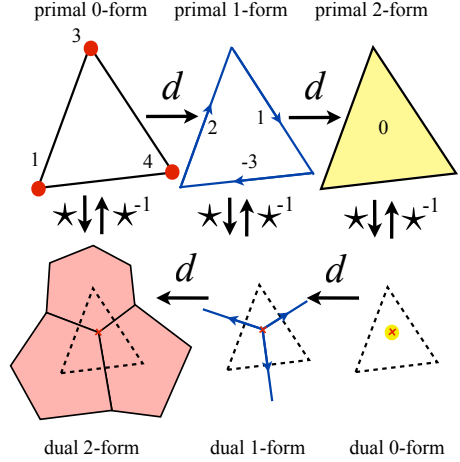}

\caption{Structure of DEC and operators for a single 2-D simplex.\label{fig:Structure-of-DEC} }
\end{centering}
\end{figure}

The formalism used to develop our discrete variational principle
is that of discrete exterior calculus (DEC). A brief overview of the basic elements 
is given here, with details found in Refs.~\onlinecite{Desbrun:2005:DDF:1198555.1198666,Desbrun:2005p5846,Elcott:2005:BYO:1198555.1198667,Bossavit:1998:Book}.
The starting point for DEC is a discrete manifold. In the simplest
case, this is a simplicial complex, essentially a collection of simplices
(lines, triangles, tetrahedra in 1-D, 2-D and 3-D respectively) embedded
in n-dimensional space. For example, a region in 3-D space discretized
using a tetrahedral mesh. The structure of differential forms in DEC
is illustrated in Figure~\ref{fig:Structure-of-DEC}, with k-forms
located on k-simplices (0-forms on vertices, 1-forms on edges etc.).
The exterior derivative operator, $d$, that takes a k-form to a (k+1)-form,
is defined so as to exactly satisfy Stoke's theorem $\int_{\chi}d\alpha=\int_{\partial\chi}\alpha$.
Importantly, with this definition, $d$ is purely topological and
$d\left(d\alpha\right)=0$. %

%%%%%%%%%%%%%%%%%%%%%%%%%%

%%%%%%%%%%%%%%%%%%%%%%%%%%

For operations involving the metric, it is necessary to define a dual
mesh, formed in this work by connecting the circumcenters of each
n-simplex (circumcentric dual). The discrete Hodge-star operator takes
a k-form on the primal mesh to an (n-k)-form on the dual mesh, see
Figure~\ref{fig:Structure-of-DEC}. The Hodge-star we use is simply
a diagonal matrix, more complex operators can give higher order accurate
theories. DEC is very well suited to the analysis of electromagnetism, in that replacing
continuous operators and forms with their discrete counterparts gives
a variational integrator with very nice properties. In fact, the very popular Yee 
staggered mesh \cite{Yee66numericalsolution} 
is simply an application of DEC on a cubic mesh \cite{Stern:2008p5799}.

The interpolation of fields to continuous space is achieved with Whitney
forms \cite{Whitney1957,Desbrun:2005:DDF:1198555.1198666,Bossavit:1998:Book}, 
which associate an interpolation k-form
to each discrete k-simplex. Using a first order scheme, Whitney 0-forms
are simply familiar {}``hat-functions,'' defined as $\varphi_{i}\left(\bm{x}\right)=1$
at vertex $i$, $\varphi_{i}\left(\bm{x}\right)=0$ at all
other vertices, with linear dependence in the neighborhood of vertex
$i$. Higher degree Whitney forms are a generalization of this. For instance,
the Whitney 1-form for the edge between node $i$ and node $j$ is
simply $\varphi_{ij}=\varphi_{i}d\varphi_{j}-\varphi_{j}d\varphi_{i},$
which is equal to $\varphi_{i}\nabla\varphi_{j}-\varphi_{j}\nabla\varphi_{i}$
if working in Euclidian space.

\begin{figure}
\begin{centering}
\includegraphics{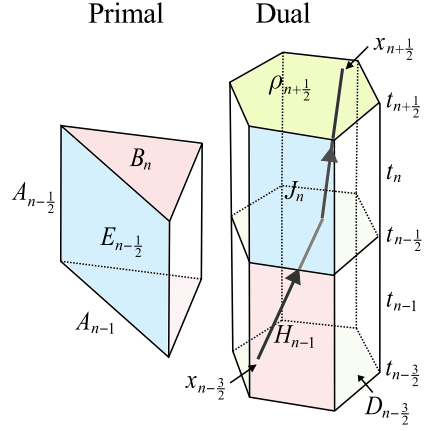}
\caption{Prismal primal and dual cells, shown in two spatial dimensions for
clarity. $A$ is a primal 1-form, $B$ and $E$ are primal 2-forms,
$D$ and $H$ are dual 2-forms, and $\star\mathcal{J}$ ($J$ and
$\rho$) is a dual 2-form. The structure of these forms in 3 spatial dimensions
is outlined in the text. Also shown is a sample particle track.\label{fig:Spacetime mesh}}

\end{centering}
\end{figure}

The discrete manifold we use in our discretization is simplex prismal
in 3+1 or 2+1 dimensions; that is, tetrahedrons or triangles projected
through time as illustrated in Figure~\ref{fig:Spacetime mesh}. 
The simpler case of an integrator on a structured cubic mesh could
also easily be derived (not done here). Since the manifold
is a direct product of a time discretization with a spatial mesh,
we can split operators into time and space components. This allows
for a simpler implementation of Maxwells equations in terms of familiar
$E$, $B$, $D$ and $H$ forms rather than the full Maxwell field
tensor, $F=dA$. It is also convenient to split $A$ into a purely
spatial 1-form (analogous to vector potential, $\bm{A}$)
and a space-time component that can be thought of as a spatial 0-form
(analogous to the scalar potential, $\phi$). These we denote by $A_{n}^{ij}$
and $A_{n+\nicefrac{1}{2}}^{i}$ respectively, due to their location
on the space-time manifold. Additionally, we split the current dual
3-form into a space-time component $J$ (spatial primal 1-form or
dual 2-form) and a purely spatial component $\rho$ (spatial primal
0-form or dual 3-form), see Figure~\ref{fig:Spacetime mesh}. Field
equations of motion as derived from the action principle are given
below.
%%%%%%%%%%%%%%%%%%%%%%%%%%

%%%%%%%%%%%%%%%%%%%%%%%%%%
In analogy with the continuous case, symmetry of a discrete action
under the transformation $A\rightarrow A+df$ (where $A$ and $f$
are discrete forms) will guarantee exact preservation of the discrete
Gauss's law for all time. The method for achieving this gauge symmetry
was motivated by Eastwood's current conserving scheme \cite{Eastwood:1991p6969}
and its recent generalization to an unstructured mesh \cite{CAMPOSPINTO:2009:HAL-00311429:3}.
This relies on integration of the particle trajectories through time
in the calculation of the discrete current. With this idea, our discrete
space-time Vlasov-Maxwell action, which is a discrete approximation
of Eq.~\eqref{eq:Continuous action}, is
\begin{align}
\mathcal{S}_{d} & =\sum_{n=0}^{N}\left\{ \sum_{\varepsilon_{s}}-\frac{1}{2}dA\wedge\star dA+h\sum_{p}\left[\frac{1}{2}m_{p}\left|\frac{\bm{x}_{n+\nicefrac{1}{2}}^{p}-\bm{x}_{n-\nicefrac{1}{2}}^{p}}{h}\right|^{2}\right.\right.\nonumber \\
 & +q_{p}\left(\frac{\bm{x}_{n+\nicefrac{1}{2}}^{p}-\bm{x}_{n-\nicefrac{1}{2}}^{p}}{h}\right) \cdot \int_{t_{n-\nicefrac{1}{2}}}^{t_{n+\nicefrac{1}{2}}}\frac{dt}{h}\left(\sum_{ij\,\epsilon\,\sigma_{1}}A_{n}^{ij}\,\bm{\varphi}_{\sigma_{ij}}\left(\bm{x}^{p}\left(t\right)\right)\right)\nonumber \\
 & -\left.\left.\frac{q_{p}}{h}\sum_{i\,\epsilon\,\sigma_{0}}A_{n-\nicefrac{1}{2}}^{i}\,\varphi_{i}\left(\bm{x}_{n-\nicefrac{1}{2}}^{p}\right)\right]\right\} .\label{eq:Full discrete action}\end{align}
Here, $h$ is the time-step, $n$ is the time index and
$p$ the particle index. $\sum_{\varepsilon_{s}}$denotes the spatial
sum of the volume form $-\frac{1}{2}dA\wedge\star dA$, and $\sum_{ij\,\epsilon\,\sigma_{1}}$
and $\sum_{i\,\epsilon\,\sigma_{0}}$ denote the sum over edges and
vertices respectively. $\bm{\varphi}_{\sigma_{ij}}$ is the
Whitney 1-form associated to edge $ij$. The particle path $\bm{x}^{p}\left(t\right)$
is taken to be linear with constant velocity $\bm{v}_n^p$ between $\bm{x}_{n-\nicefrac{1}{2}}^{p}$
and $\bm{x}_{n+\nicefrac{1}{2}}^{p}$. As is standard in variational
integrators, the particle Lagrangian is designed to approximate $\int_{t_{n-1/2}}^{t_{n+1/2}}dt\, L_{c}^{p}$
where $L_{c}^{p}$ is the continuous Lagrangian. The present case is that
of a single particle in the discrete electromagnetic field.

The field part of Eq.~\eqref{eq:Full discrete action} is obviously
gauge invariant since $d^{2}=0$. Since the particle part is linear
in $A$, gauge invariance can be seen by substituting $A=df$ and
showing that this only gives contributions from the endpoints. This
is straightforward using $d_{c}\left(\left(\alpha\right)_{interp}\right)=\left(d_{d}\alpha\right)_{interp}$
for a 0-form $\alpha$, where $d_{c}$ and $d_{d}$ are the continuous
and discrete exterior derivatives and $\left(\right)_{interp}$ signifies
Whitney interpolation of the form. 

Field equations arise from variation of the discrete action with respect
to the potential, $A$, yielding  (see Ref.~\onlinecite{Stern:2008p5799})
\begin{equation}
d\star dA=\mathcal{J}.\label{eq:EM eqns}\end{equation}
Due to the tensor product nature of the discrete manifold this is
equivalent to 
\begin{align}
d_{s} & E_{n+\nicefrac{1}{2}}+\frac{B_{n+1}-B_{n}}{h}=0\label{eq:Maxwell 1}\\
d_{s}H_{n} & -\frac{D_{n+\nicefrac{1}{2}}-D_{n-\nicefrac{1}{2}}}{h}=\sum_{p}q_{p}\,\bm{v}_{n}^p \cdot \int_{t_{n-\nicefrac{1}{2}}}^{t_{n+\nicefrac{1}{2}}}\frac{dt}{h}\,\bm{\varphi}_{\sigma_{ij}}\mid_{\bm{x}^{p}\left(t\right)},\label{eq:Maxwell 2}
\end{align}
and
\begin{align}
 & d_{s}D_{n-\nicefrac{1}{2}}=\sum_{p} q_{p}\,\varphi_{i}\mid_{\bm{x}_{n-\nicefrac{1}{2}}^{p}}\label{eq:Gauss's law}\\
 & d_{s}B=0.\label{eq:divB constraint}
 \end{align}
Here $E_{n+\nicefrac{1}{2}}=-\frac{1}{h}\left(A_{n+1}-A_{n}\right)-d_{s}A_{n+\nicefrac{1}{2}}$
is a spatial primal 1-form, $B_{n}=d_{s}A_{n}$ is a spatial
primal 2-form, $D=\star_{s}E$ and $H=\star_{s}B$, with the subscript
$s$ indicating the spatial part of a DEC operator. Note that Eqns.~\eqref{eq:Gauss's law}
and \eqref{eq:divB constraint} are constraints and need only be applied
as initial conditions. The particle equations of motion are derived
from variations of Eq.~\eqref{eq:Full discrete action} with respect
to $\bm{x}_{n-\nicefrac{1}{2}}^{p}$. This leads to the particle
equations of motion,\begin{align}
\frac{1}{h^{2}} & \left(\bm{x}_{n+\nicefrac{1}{2}}^{p}-2\bm{x}_{n-\nicefrac{1}{2}}^{p}+\bm{x}_{n-\nicefrac{3}{2}}^{p}\right)\nonumber \\
 & =\frac{q_{p}}{m_{p}}\left(\tilde{\bm{E}}_{n-\nicefrac{1}{2}}^{p}+\frac{1}{2}\bm{v}_{n}^{p}\times\tilde{\bm{B}}_{n}^{p}+\frac{1}{2}\bm{v}_{n-1}^{p}\times\tilde{\bm{B}}_{n-1}^{p}\right),\label{eq:Particle eqn of motion}\end{align}
where \begin{align}
 & \tilde{\bm{E}}_{n-\nicefrac{1}{2}}^{p}=\left(E_{n-\nicefrac{1}{2}}\right)_{interp}\mid_{\bm{x}_{n-\nicefrac{1}{2}}^{p}}\label{eq:Enmh interpolation}\\
 & \tilde{\bm{B}}_{n}^{p}=\int_{t_{n-\nicefrac{1}{2}}}^{t_{n+\nicefrac{1}{2}}}\frac{dt}{h}\left(\frac{t_{n+\nicefrac{1}{2}}-t}{h}\right)\left(B_{n}^{p}\,\right)_{interp}\mid_{\bm{x}^{p}\left(t\right)}\label{eq:Bn interpolation}\\
 & \tilde{\bm{B}}_{n-1}^{p}=\int_{t_{n-\nicefrac{3}{2}}}^{t_{n-\nicefrac{1}{2}}}\frac{dt}{h}\left(\frac{t-t_{n-\nicefrac{3}{2}}}{h}\right)\left(B_{n-1}^{p}\right)_{interp}\mid_{\bm{x}^{p}\left(t\right)}\label{eq:Bnm1 interpolation}
\end{align}
Since particle trajectories are linear and fields, $ $$\left(E_{n-\nicefrac{1}{2}}\right)_{interp}$
and $\left(B_{n}^{p}\,\right)_{interp}$, are piecewise polynomial,
time integrals can be performed exactly using Gaussian quadrature\cite{CAMPOSPINTO:2009:HAL-00311429:3}. Because of the $\tilde{\bm{B}}_{n}^{p}$
term, the algorithm is implicit. However, if a quasi-particle stays in the same cell, as 
is the case for the majority of time-steps, Eq.~\eqref{eq:Particle eqn of motion}
can be easily solved without resorting to an iterative scheme.

While variational formulations have been used for PIC methods in the
past\cite{BirdsallLangdon}, this is, to our knowledge, the
first PIC scheme to use a full space-time variational principle. As
a consequence, the algorithm as a whole is multi-symplectic\cite{Marsden:1998p5813,Stern:2008p5799}, an important geometrical
property proven to have profound consequences for the integration
of systems of ordinary differential equations \cite{Reich:1999:BEA:333858.333895,Marsden:2001varint}.
Though presented from a somewhat different standpoint, the algorithm
is similar to those in Ref.~\onlinecite{CAMPOSPINTO:2009:HAL-00311429:3}.
The crucial difference is that our particle mover is constrained to
a particular form by the discrete action, which is necessary for a
fully multi-symplectic method. 

Using the ideas in Ref.~\onlinecite{Elcott:2005:BYO:1198555.1198667}, the unstructured
Maxwell solver is very simple to implement. Field advancement is governed by Eqs.~\eqref{eq:Maxwell 1}
and \eqref{eq:Maxwell 2}, and simply involves sparse matrix multiplication.
As a test case, we have implemented a 2-D version of variational PIC in \textsc{Matlab}, with
a magnetic field directed out of the plane. In this case, the equations
of motion and definitions of $E$, $B$, $D$, and $H$ are exactly
the same as the 3-D case. Particle advancement is implemented by first
assuming the particles stay in the same cell. In the case where this
is not true and an implicit solver is needed, the current contribution
to the grid is calculated at the same time as particle advancement.
As a consequence, the extra computational expense over an explicit particle
pusher scheme is minimal.

\begin{figure}
\begin{centering}

\includegraphics{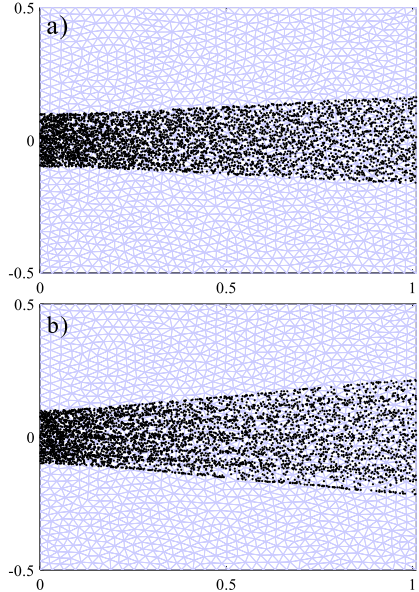}
\caption{Electron beam particle 
distribution accelerated by external potential. Evolved in time using: (a) 
symplectic PIC (current conserving) (b) non current conserving algorithm producing spurious bunching
of the charge into lines due to violations of Gauss's law.\label{fig:Oscillating particle}}

\end{centering}
\end{figure}

Investigations are ongoing into the numerical properties of variational
PIC, with special focus on the importance of the multi-symplectic nature of the
algorithm. Here we give a brief numerical example, motivated by 
Refs.~\onlinecite{Bartheme_PICcurrent,CAMPOSPINTO:2009:HAL-00311429:3}, 
to illustrate the importance of numerical current conservation. 
On a triangular mesh, a beam of electrons is accelerated by an external 
voltage (from left to right) calculated to satisfy  the Child-Langmuir law. 
Figure~\ref{fig:Oscillating particle}(a) shows the particle distribution at $t=40$ 
using symplectic PIC, while Figure~\ref{fig:Oscillating particle}(b) illustrates 
the distribution for the same initial conditions, advanced using an integrator that 
does not conserve current. Local violations in Gauss's law caused by lack of current conservation
are manifested through unphysical bunching of the charge into lines of higher 
density, as evident in Figure~\ref{fig:Oscillating particle}(b). The beam also
widens more than in the current conserving case, showing that there is an 
overestimation of the  average self electric field.

%%%%%%%%%%%%%%%%%%%%%%
%%%%%%%%%%%%%%%%%%%%%%%%

Our discretization of the variational principle [Eq.~\eqref{eq:Full discrete action}]
is relatively arbitrary. As an avenue for future work it could be
interesting to explore different discretizations of the same system
(for instance, different particle pushers, particle shape factors
or field Hodge-star operators) to better understand some advantages
of a fully multi-symplectic scheme. Another approach would be to design
a fully implicit variational PIC integrator. At the cost of complexity,
implicit PIC schemes circumvent many of the numerical instabilities
inherent in explicit PIC \cite{BirdsallLangdon,Lapenta2011} and allow
larger time-steps.

The methods presented above will allow relatively simple generalizations to
more complex mesh schemes. One such idea would be an Asynchronous
Variational Integrator, in which each grid cell and particle could
be advanced with a different time-step set by its own Courant condition
\cite{Lew2003_AVIs,Stern:2008p5799}.  Time savings can be substantial on 
highly irregular meshes. As a further generalization of this type
of idea, a 4-D simplicial complex could be used in a completely covariant
general relativistic PIC code, which would have many astrophysical
applications. The DEC and variational formalisms could be very important
in formulating methods for field advancement and current conservation
in these complex systems.

Perhaps one of the most exciting areas of future research is in the
design of geometric PIC algorithms for more complex field theories,
in particular gyrokinetics \cite{Qin:2007p5801}. Being non-local,
gyrokinetics presents a great challenge in algorithm design if one
is to respect important geometrical properties of the system. The
question of how to ensure current conservation is answered very cleanly
by the realization that it is simply the requirement that a discrete
variational principle be electromagnetically gauge invariant. 

This research was supported by the U.S.~Department 
of Energy under Contract No.~AC02-09CH11466.


\begin{thebibliography}{24}%
\makeatletter
\providecommand \@ifxundefined [1]{%
 \@ifx{#1\undefined}
}%
\providecommand \@ifnum [1]{%
 \ifnum #1\expandafter \@firstoftwo
 \else \expandafter \@secondoftwo
 \fi
}%
\providecommand \@ifx [1]{%
 \ifx #1\expandafter \@firstoftwo
 \else \expandafter \@secondoftwo
 \fi
}%
\providecommand \natexlab [1]{#1}%
\providecommand \enquote  [1]{``#1''}%
\providecommand \bibnamefont  [1]{#1}%
\providecommand \bibfnamefont [1]{#1}%
\providecommand \citenamefont [1]{#1}%
\providecommand \href@noop [0]{\@secondoftwo}%
\providecommand \href [0]{\begingroup \@sanitize@url \@href}%
\providecommand \@href[1]{\@@startlink{#1}\@@href}%
\providecommand \@@href[1]{\endgroup#1\@@endlink}%
\providecommand \@sanitize@url [0]{\catcode `\\12\catcode `\$12\catcode
  `\&12\catcode `\#12\catcode `\^12\catcode `\_12\catcode `\%12\relax}%
\providecommand \@@startlink[1]{}%
\providecommand \@@endlink[0]{}%
\providecommand \url  [0]{\begingroup\@sanitize@url \@url }%
\providecommand \@url [1]{\endgroup\@href {#1}{\urlprefix }}%
\providecommand \urlprefix  [0]{URL }%
\providecommand \Eprint [0]{\href }%
\providecommand \doibase [0]{http://dx.doi.org/}%
\providecommand \selectlanguage [0]{\@gobble}%
\providecommand \bibinfo  [0]{\@secondoftwo}%
\providecommand \bibfield  [0]{\@secondoftwo}%
\providecommand \translation [1]{[#1]}%
\providecommand \BibitemOpen [0]{}%
\providecommand \bibitemStop [0]{}%
\providecommand \bibitemNoStop [0]{.\EOS\space}%
\providecommand \EOS [0]{\spacefactor3000\relax}%
\providecommand \BibitemShut  [1]{\csname bibitem#1\endcsname}%
\let\auto@bib@innerbib\@empty
%</preamble>
\bibitem [{\citenamefont {Desbrun}\ \emph
  {et~al.}(2005{\natexlab{a}})\citenamefont {Desbrun}, \citenamefont {Hirani},
  \citenamefont {Leok},\ and\ \citenamefont {Marsden}}]{Desbrun:2005p5846}%
  \BibitemOpen
  \bibfield  {author} {\bibinfo {author} {\bibfnamefont {M.}~\bibnamefont
  {Desbrun}}, \bibinfo {author} {\bibfnamefont {A.~N.}\ \bibnamefont {Hirani}},
  \bibinfo {author} {\bibfnamefont {M.}~\bibnamefont {Leok}}, \ and\ \bibinfo
  {author} {\bibfnamefont {J.~E.}\ \bibnamefont {Marsden}},\ }\href@noop {} {}
  (\bibinfo {year} {2005}{\natexlab{a}}),\ \Eprint
  {http://arxiv.org/abs/math/0508341} {arXiv:math/0508341} \BibitemShut
  {NoStop}%
\bibitem [{\citenamefont {Rosner}(2010)}]{RRosner_Computing2010}%
  \BibitemOpen
  \bibfield  {author} {\bibinfo {author} {\bibfnamefont {R.}~\bibnamefont
  {Rosner}} (\bibinfo {collaboration} {DoE Advanced Scientific Computing
  Advisory Committee Report}),\ }\href@noop {} {} (\bibinfo {year}
  {2010})\BibitemShut {NoStop}%
\bibitem [{\citenamefont {Veselov}(1988)}]{springerlink:10.1007/BF01077598}%
  \BibitemOpen
  \bibfield  {author} {\bibinfo {author} {\bibfnamefont {A.~P.}\ \bibnamefont
  {Veselov}},\ }\href@noop {} {\bibfield  {journal} {\bibinfo  {journal}
  {Functional Analysis and Its Applications}\ }\textbf {\bibinfo {volume}
  {22}},\ \bibinfo {pages} {83} (\bibinfo {year} {1988})}\BibitemShut {NoStop}%
\bibitem [{\citenamefont {Marsden}\ and\ \citenamefont
  {West}(2001)}]{Marsden:2001varint}%
  \BibitemOpen
  \bibfield  {author} {\bibinfo {author} {\bibfnamefont {J.~E.}\ \bibnamefont
  {Marsden}}\ and\ \bibinfo {author} {\bibfnamefont {M.}~\bibnamefont {West}},\
  }\href@noop {} {\bibfield  {journal} {\bibinfo  {journal} {Acta Numerica}\
  }\textbf {\bibinfo {volume} {10}},\ \bibinfo {pages} {357} (\bibinfo {year}
  {2001})}\BibitemShut {NoStop}%
\bibitem [{\citenamefont {Qin}\ and\ \citenamefont
  {Guan}(2008)}]{Qin:2008p5812}%
  \BibitemOpen
  \bibfield  {author} {\bibinfo {author} {\bibfnamefont {H.}~\bibnamefont
  {Qin}}\ and\ \bibinfo {author} {\bibfnamefont {X.}~\bibnamefont {Guan}},\
  }\href@noop {} {\bibfield  {journal} {\bibinfo  {journal} {Phys. Rev. Lett.}\
  }\textbf {\bibinfo {volume} {100}},\ \bibinfo {pages} {035006} (\bibinfo
  {year} {2008})}\BibitemShut {NoStop}%
\bibitem [{\citenamefont {Bochev}\ and\ \citenamefont
  {Hyman}(2006)}]{Hyman-2005-principles}%
  \BibitemOpen
  \bibfield  {author} {\bibinfo {author} {\bibfnamefont {P.~B.}\ \bibnamefont
  {Bochev}}\ and\ \bibinfo {author} {\bibfnamefont {J.~M.}\ \bibnamefont
  {Hyman}},\ }\href@noop {} {\bibfield  {journal} {\bibinfo  {journal} {IMA
  vols. math. and appl.}\ }\textbf {\bibinfo {volume} {142}},\ \bibinfo {pages}
  {89} (\bibinfo {year} {2006})}\BibitemShut {NoStop}%
\bibitem [{\citenamefont {Bossavit}(1998)}]{Bossavit:1998:Book}%
  \BibitemOpen
  \bibfield  {author} {\bibinfo {author} {\bibfnamefont {A.}~\bibnamefont
  {Bossavit}},\ }\href@noop {} {\emph {\bibinfo {title} {Computational
  Electromagnetism}}}\ (\bibinfo  {publisher} {Academic Press Inc.},\ \bibinfo
  {address} {San Diego CA},\ \bibinfo {year} {1998})\BibitemShut {NoStop}%
\bibitem [{\citenamefont {Lew}\ \emph {et~al.}(2003)\citenamefont {Lew},
  \citenamefont {Marsden}, \citenamefont {Ortiz},\ and\ \citenamefont
  {West}}]{Lew2003_AVIs}%
  \BibitemOpen
  \bibfield  {author} {\bibinfo {author} {\bibfnamefont {A.}~\bibnamefont
  {Lew}}, \bibinfo {author} {\bibfnamefont {J.~E.}\ \bibnamefont {Marsden}},
  \bibinfo {author} {\bibfnamefont {M.}~\bibnamefont {Ortiz}}, \ and\ \bibinfo
  {author} {\bibfnamefont {M.}~\bibnamefont {West}},\ }\href@noop {} {\bibfield
   {journal} {\bibinfo  {journal} {Arch. Rational Mechanics and Analysis}\
  }\textbf {\bibinfo {volume} {167}},\ \bibinfo {pages} {85} (\bibinfo {year}
  {2003})}\BibitemShut {NoStop}%
\bibitem [{\citenamefont {Stern}\ \emph {et~al.}(2007)\citenamefont {Stern},
  \citenamefont {Tong}, \citenamefont {Desbrun},\ and\ \citenamefont
  {Marsden}}]{Stern:2008p5799}%
  \BibitemOpen
  \bibfield  {author} {\bibinfo {author} {\bibfnamefont {A.}~\bibnamefont
  {Stern}}, \bibinfo {author} {\bibfnamefont {Y.}~\bibnamefont {Tong}},
  \bibinfo {author} {\bibfnamefont {M.}~\bibnamefont {Desbrun}}, \ and\
  \bibinfo {author} {\bibfnamefont {J.~E.}\ \bibnamefont {Marsden}},\
  }\href@noop {} {} (\bibinfo {year} {2007}),\ \Eprint
  {http://arxiv.org/abs/0707.4470v3 [math.NA]} {arXiv:0707.4470v3 [math.NA]}
  \BibitemShut {NoStop}%
\bibitem [{\citenamefont {Pavlov}\ \emph {et~al.}(2011)\citenamefont {Pavlov},
  \citenamefont {Mullen}, \citenamefont {Tong}, \citenamefont {Kanso},
  \citenamefont {Marsden},\ and\ \citenamefont {Desbrun}}]{Pavlov2011443}%
  \BibitemOpen
  \bibfield  {author} {\bibinfo {author} {\bibfnamefont {D.}~\bibnamefont
  {Pavlov}}, \bibinfo {author} {\bibfnamefont {P.}~\bibnamefont {Mullen}},
  \bibinfo {author} {\bibfnamefont {Y.}~\bibnamefont {Tong}}, \bibinfo {author}
  {\bibfnamefont {E.}~\bibnamefont {Kanso}}, \bibinfo {author} {\bibfnamefont
  {J.}~\bibnamefont {Marsden}}, \ and\ \bibinfo {author} {\bibfnamefont
  {M.}~\bibnamefont {Desbrun}},\ }\href@noop {} {\bibfield  {journal} {\bibinfo
   {journal} {Physica D}\ }\textbf {\bibinfo {volume} {240}},\ \bibinfo {pages}
  {443 } (\bibinfo {year} {2011})}\BibitemShut {NoStop}%
\bibitem [{\citenamefont {Gawlik}\ \emph {et~al.}(2011)\citenamefont {Gawlik},
  \citenamefont {Mullen}, \citenamefont {Pavlov}, \citenamefont {Marsden},\
  and\ \citenamefont {Desbrun}}]{Gawlik:2010p5814}%
  \BibitemOpen
  \bibfield  {author} {\bibinfo {author} {\bibfnamefont {E.}~\bibnamefont
  {Gawlik}}, \bibinfo {author} {\bibfnamefont {P.}~\bibnamefont {Mullen}},
  \bibinfo {author} {\bibfnamefont {D.}~\bibnamefont {Pavlov}}, \bibinfo
  {author} {\bibfnamefont {J.~E.}\ \bibnamefont {Marsden}}, \ and\ \bibinfo
  {author} {\bibfnamefont {M.}~\bibnamefont {Desbrun}},\ }\href@noop {}
  {\bibfield  {journal} {\bibinfo  {journal} {Physica D}\ }\textbf {\bibinfo
  {volume} {240}},\ \bibinfo {pages} {1724} (\bibinfo {year}
  {2011})}\BibitemShut {NoStop}%
\bibitem [{\citenamefont {Qin}\ \emph {et~al.}(2007)\citenamefont {Qin},
  \citenamefont {Cohen}, \citenamefont {Nevins},\ and\ \citenamefont
  {Xu}}]{Qin:2007p5801}%
  \BibitemOpen
  \bibfield  {author} {\bibinfo {author} {\bibfnamefont {H.}~\bibnamefont
  {Qin}}, \bibinfo {author} {\bibfnamefont {R.~H.}\ \bibnamefont {Cohen}},
  \bibinfo {author} {\bibfnamefont {W.~M.}\ \bibnamefont {Nevins}}, \ and\
  \bibinfo {author} {\bibfnamefont {X.~Q.}\ \bibnamefont {Xu}},\ }\href@noop {}
  {\bibfield  {journal} {\bibinfo  {journal} {Phys. Plasmas}\ }\textbf
  {\bibinfo {volume} {14}},\ \bibinfo {pages} {056110} (\bibinfo {year}
  {2007})}\BibitemShut {NoStop}%
\bibitem [{\citenamefont {Low}(1958)}]{Low:1958p5805}%
  \BibitemOpen
  \bibfield  {author} {\bibinfo {author} {\bibfnamefont {F.}~\bibnamefont
  {Low}},\ }\href@noop {} {\bibfield  {journal} {\bibinfo  {journal} {Proc. R.
  Soc. A}\ ,\ \bibinfo {pages} {282}} (\bibinfo {year} {1958})}\BibitemShut
  {NoStop}%
\bibitem [{\citenamefont {Desbrun}\ \emph
  {et~al.}(2005{\natexlab{b}})\citenamefont {Desbrun}, \citenamefont {Kanso},\
  and\ \citenamefont {Tong}}]{Desbrun:2005:DDF:1198555.1198666}%
  \BibitemOpen
  \bibfield  {author} {\bibinfo {author} {\bibfnamefont {M.}~\bibnamefont
  {Desbrun}}, \bibinfo {author} {\bibfnamefont {E.}~\bibnamefont {Kanso}}, \
  and\ \bibinfo {author} {\bibfnamefont {Y.}~\bibnamefont {Tong}},\ }in\
  \href@noop {} {\emph {\bibinfo {booktitle} {ACM SIGGRAPH 2005 Courses}}},\
  \bibinfo {series and number} {SIGGRAPH '05}\ (\bibinfo  {publisher} {ACM},\
  \bibinfo {address} {New York, NY, USA},\ \bibinfo {year} {2005})\BibitemShut
  {NoStop}%
\bibitem [{\citenamefont {Elcott}\ and\ \citenamefont
  {Schr\"{o}der}(2005)}]{Elcott:2005:BYO:1198555.1198667}%
  \BibitemOpen
  \bibfield  {author} {\bibinfo {author} {\bibfnamefont {S.}~\bibnamefont
  {Elcott}}\ and\ \bibinfo {author} {\bibfnamefont {P.}~\bibnamefont
  {Schr\"{o}der}},\ }in\ \href@noop {} {\emph {\bibinfo {booktitle} {ACM
  SIGGRAPH 2005 Courses}}},\ \bibinfo {series and number} {SIGGRAPH '05}\
  (\bibinfo  {publisher} {ACM},\ \bibinfo {address} {New York, NY, USA},\
  \bibinfo {year} {2005})\BibitemShut {NoStop}%
\bibitem [{\citenamefont {Yee}(1966)}]{Yee66numericalsolution}%
  \BibitemOpen
  \bibfield  {author} {\bibinfo {author} {\bibfnamefont {K.~S.}\ \bibnamefont
  {Yee}},\ }\href@noop {} {\bibfield  {journal} {\bibinfo  {journal} {IEEE
  Trans. Ant. Prop.}\ ,\ \bibinfo {pages} {302}} (\bibinfo {year}
  {1966})}\BibitemShut {NoStop}%
\bibitem [{\citenamefont {Whitney}(1957)}]{Whitney1957}%
  \BibitemOpen
  \bibfield  {author} {\bibinfo {author} {\bibfnamefont {H.}~\bibnamefont
  {Whitney}},\ }\href@noop {} {\emph {\bibinfo {title} {Geometric integration
  theory}}}\ (\bibinfo  {publisher} {Princeton university press},\ \bibinfo
  {year} {1957})\BibitemShut {NoStop}%
\bibitem [{\citenamefont {Eastwood}(1991)}]{Eastwood:1991p6969}%
  \BibitemOpen
  \bibfield  {author} {\bibinfo {author} {\bibfnamefont {J.}~\bibnamefont
  {Eastwood}},\ }\href@noop {} {\bibfield  {journal} {\bibinfo  {journal}
  {Comp. phys. comm.}\ }\textbf {\bibinfo {volume} {64}},\ \bibinfo {pages}
  {252} (\bibinfo {year} {1991})}\BibitemShut {NoStop}%
\bibitem [{\citenamefont {Campos-Pinto}\ \emph {et~al.}(2009)\citenamefont
  {Campos-Pinto}, \citenamefont {Jund}, \citenamefont {Salmon},\ and\
  \citenamefont {Sonnendr{\"u}cker}}]{CAMPOSPINTO:2009:HAL-00311429:3}%
  \BibitemOpen
  \bibfield  {author} {\bibinfo {author} {\bibfnamefont {M.}~\bibnamefont
  {Campos-Pinto}}, \bibinfo {author} {\bibfnamefont {S.}~\bibnamefont {Jund}},
  \bibinfo {author} {\bibfnamefont {S.}~\bibnamefont {Salmon}}, \ and\ \bibinfo
  {author} {\bibfnamefont {E.}~\bibnamefont {Sonnendr{\"u}cker}} (\bibinfo
  {collaboration} {Projet Calvi (INRIA), ANR Houpic.}),\ }\href
  {http://hal.archives-ouvertes.fr/hal-00311429} {} (\bibinfo {year}
  {2009})\BibitemShut {NoStop}%
\bibitem [{\citenamefont {Birdsall}\ and\ \citenamefont
  {Langdon}(1991)}]{BirdsallLangdon}%
  \BibitemOpen
  \bibfield  {author} {\bibinfo {author} {\bibfnamefont {C.~K.}\ \bibnamefont
  {Birdsall}}\ and\ \bibinfo {author} {\bibfnamefont {A.~B.}\ \bibnamefont
  {Langdon}},\ }\href@noop {} {\emph {\bibinfo {title} {Plasma Physics via
  Computer Simulation}}}\ (\bibinfo  {publisher} {IOP Publishing Ltd.},\
  \bibinfo {year} {1991})\BibitemShut {NoStop}%
\bibitem [{\citenamefont {Marsden}\ \emph {et~al.}(1998)\citenamefont
  {Marsden}, \citenamefont {Patrick},\ and\ \citenamefont
  {Shkoller}}]{Marsden:1998p5813}%
  \BibitemOpen
  \bibfield  {author} {\bibinfo {author} {\bibfnamefont {J.}~\bibnamefont
  {Marsden}}, \bibinfo {author} {\bibfnamefont {G.~W.}\ \bibnamefont
  {Patrick}}, \ and\ \bibinfo {author} {\bibfnamefont {S.}~\bibnamefont
  {Shkoller}},\ }\href@noop {} {\bibfield  {journal} {\bibinfo  {journal}
  {Comm. in Math. Phys.}\ }\textbf {\bibinfo {volume} {199}},\ \bibinfo {pages}
  {351} (\bibinfo {year} {1998})}\BibitemShut {NoStop}%
\bibitem [{\citenamefont {Reich}(1999)}]{Reich:1999:BEA:333858.333895}%
  \BibitemOpen
  \bibfield  {author} {\bibinfo {author} {\bibfnamefont {S.}~\bibnamefont
  {Reich}},\ }\href@noop {} {\bibfield  {journal} {\bibinfo  {journal} {SIAM J.
  Numer. Anal.}\ }\textbf {\bibinfo {volume} {36}},\ \bibinfo {pages} {1549}
  (\bibinfo {year} {1999})}\BibitemShut {NoStop}%
\bibitem [{\citenamefont {Barthelm\'{e}}\ and\ \citenamefont
  {Parzani}(2005)}]{Bartheme_PICcurrent}%
  \BibitemOpen
  \bibfield  {author} {\bibinfo {author} {\bibfnamefont {R.}~\bibnamefont
  {Barthelm\'{e}}}\ and\ \bibinfo {author} {\bibfnamefont {C.}~\bibnamefont
  {Parzani}},\ }\href@noop {} {\bibfield  {journal} {\bibinfo  {journal} {IRMA
  Lect. Math. Theor. Phys.}\ }\textbf {\bibinfo {volume} {7}},\ \bibinfo
  {pages} {7} (\bibinfo {year} {2005})}\BibitemShut {NoStop}%
\bibitem [{\citenamefont {Lapenta}\ and\ \citenamefont
  {Markidis}(2011)}]{Lapenta2011}%
  \BibitemOpen
  \bibfield  {author} {\bibinfo {author} {\bibfnamefont {G.}~\bibnamefont
  {Lapenta}}\ and\ \bibinfo {author} {\bibfnamefont {S.}~\bibnamefont
  {Markidis}},\ }\href@noop {} {\bibfield  {journal} {\bibinfo  {journal}
  {Phys. Plasmas}\ }\textbf {\bibinfo {volume} {18}},\ \bibinfo {pages}
  {072101} (\bibinfo {year} {2011})}\BibitemShut {NoStop}%
\end{thebibliography}
\end{document}